\documentclass{aa}

\input psfig
\begin{document}

   \thesaurus{01          
              ( 
                12.03.3;  
                12.07.1;  
                11.17.4;  
                13.25.2   
                13.25.3   
		)}

\title{The gravitationally lensed quasar Q2237+0305 in X-rays:
	ROSAT/HRI detection of the ``Einstein Cross"}

   \author{	
   		Joachim Wambsganss$^1$,
   		Hermann Brunner$^1$,
   		Sabine Schindler$^2$,
   		Emilio Falco$^3$,
	}

   \institute{
       	       Astrophysikalisches Institut Potsdam,
	       Sternwarte 16,
               14482 Potsdam,
               Germany;
               e-mail: {\tt jwambsganss, hbrunner@aip.de}\\
\and
		Astrophysics Research Institute, 
		Liverpool John Moores University, 
		Twelve Quays House, Egerton Wharf,
		Birkenhead L41 1LD,
		United Kingdom;
               e-mail: {\tt sas@astro.livjm.ac.uk}\\
\and
       	       Harvard-Smithsonian Center for Astrophysics,
	       60 Garden St.,
	       Cambridge, MA 02138
	       USA.
               e-mail: {\tt falco@cfa.harvard.edu}\\
              }
\authorrunning {Joachim Wambsganss et al.}
\titlerunning {ROSAT/HRI  observations of Einstein Cross Q2237+0305}
   \maketitle
\markboth
{draft printed on    \hspace{3.5cm} \today}
{draft printed on    \hspace{3.5cm} \today}

\begin{abstract}

We report the first detection of the gravitationally lensed quasar
Q2237+0305 in X-rays.  With a ROSAT/HRI exposure of 53 ksec taken in
Nov./Dec. 1997, we found a count rate of 0.006 counts per second for
the combined four images.  This corresponds to an X-ray flux of
$2.2\times10^{-13}$ erg/cm$^2$/sec and an X-ray luminosity of
$4.2\times10^{45}$erg/sec (in the ROSAT energy window 0.1-2.4 keV).
The ROSAT/HRI detector is not able to resolve spatially the four
quasar images (maximum separation 1.8 arcsec).  The analysis is based
on about 330 source photons.  The signal is consistent with no
variability, but with low significance.  
This detection is promising
in view of the upcoming X-ray missions with higher spatial/spectral
resolution and/or collecting power (Chandra X-ray Observatory, 
XMM and ASTRO-E).

\end{abstract}

\section{Introduction}

The quasar Q2237+0305 at a redshift of $z = 1.609$ is lensed by a
relatively nearby galaxy at $z_L = 0.039$ (\cite{huc85}).  It is a
quadruply-imaged case, and one of the best investigated lens systems,
both observationally and theoretically. For recent work, see
\cite{cha98,yon98} and \cite{bla98,med98}, Schmidt et al. (1998),
respectively.

Q2237+0305 was the first multiple quasar system in which microlensing
was detected (see, e.g. \cite{irw89}, \cite{cor91}, \cite{lew98}).
The analysis of well covered microlensing light curves of a quasar can
be used to uncover its size and structure (Wambsganss et al. 1990,
Wambsganss \& Paczy\'nski 1992).  A number of groups are optically
monitoring this system to measure any microlensing effects.  The
expected time delay between the four images is only of order a day
(\cite{rix92, wam94}) and hence unlikely to be determined from optical
light curves.

Recently, HST observations in the UV allowed the determination of
highly accurate relative positions of the four images (Blanton et
al. 1998).  With ground-based spectrophotometry, an extended arc
comprising three of the four images was discovered (Mediavilla et
al. 1998).

Not very much is known yet about gravitationally lensed quasars in
X-rays.  The double quasar Q0957+561 was seen with HEAO-1 and with
ROSAT, and dramatic differences in the flux of image B of up to a
factor of five were observed \cite{chartas95}.  There is an X-ray
selected gravitationally lensed quasar, RX J0911.4+0551 which was
found in the ROSAT All-Sky Survey \cite{bade97}.  With $z_Q=2.8$ and
an X-ray luminosity of $L_X=4.1 \times 10^{46}$ergs/s it is a very
X-ray bright quasar.  The two bright images are separated by 0.8
arcsec.  High-resolution optical/infrared imaging revealed four images
with a maximum distance of 3.1 arcsec (\cite{bur98}, see also
Munoz et al. 1999 or
http://cfa-www.harvard.edu/castles for HST/NICMOS data obtained by
the CASTLES collaboration).

Here we present the first X-ray detection of Q2237+0305, an analysis
of a ROSAT/HRI observation.  The combined X-ray emission of the four
quasar images is clearly detected, though at a relatively low count
rate.  Due to the coarse resolution of the ROSAT/HRI the individual
images are not resolved.

\section{Observations}

We observed the quasar Q2237+0305 with the ROSAT/HRI (\cite{tr83}) for
a total exposure time of $t_{\rm ex} = 53 869.7$ seconds.  The
observations took place between November 20 and December 5, 1997.  The
Standard Analysis Software System (SASS) determined an average
background rate of 0.0032 counts/sec/arcmin$^2$ which -- multiplied by
the ``exposure time" $t_{\rm ex}$ yields an average of 171.7
background counts per square arcmin in total.

To determine the count rate of Q2237+0305, we extracted the photons in
circles of different sizes and subtracted the background, whose count
rate was determined from empty regions of considerably larger size.  A
circle with a radius of 15 arcseconds centered on the pixel with the
highest count (RA 22:40:30.21, Dec +03:21:28.7; J2000) resulted in a
total number of 361 counts.  The average number of background photons
determined from ten ``empty" circles nearby resulted in 39.8
background photons.  This leaves 321.2 source counts, which results in
a count rate of 6.0 counts per kilosecond.
A similar determination with a much larger extraction radius of 50
arcseconds (100 pixels) centered on position RA 22:40:30.0, Dec
+03:21:28.7 produced 800 counts. The average background for this size
is 466.8 counts, which results in 333.2 source counts and a count rate
of $(6.2 \pm 2.8)$ counts/ksec.

\begin{figure}[htb]
\psfig{figure=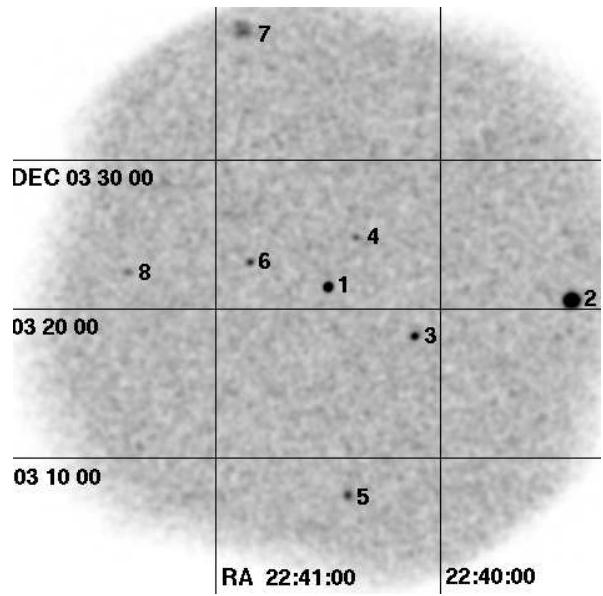            ,width=7.8cm,clip=} 
\caption{\label{fig-hri} ROSAT/HRI image of the full field 
of Q2237+0305 (center, labelled ``1"). The field size is about 1 degree.
Positions and count rates of the numbered X-ray detections are 
given in Table \ref{tab-sources}.  The data is smoothed with a Gaussian
of $\sigma=5$ arcsec.
}
\end{figure}

The complete HRI field of the exposure is shown in Figure
\ref{fig-hri}. The central source is Q2237+0305 (labelled ``1").
Table \ref{tab-sources} contains the positions and count rates of all
the sources in the field that are detected with a S/N of at least 4.0
in one of the ROSAT/HRI detection cells (squares with sizes ranging
from 12 to 120 arcsec side length).  In the table we also list the
count rates and possible identifications of these X-ray detections.
Cross-checking with the databases SIMBAD and NED, 
we
could find one other identification of an X-ray detection in our field
with a catalogued source (aside from the ``target" Q2237+0305): Source
No. 2 coincides with the G0 star BD +02 4540 (V magnitude: 9.6).  The
G5 star HD 214787 (V magnitude: 8.3) is about 50 arcsec off the
position of detection No. 5, but this is a very unlikely match, even
considering the poor accuracy for large off-axis angles.

%
%
%
%
%
%
%
%
\begin{table*}[htb]
\begin{center}
\begin{tabular}{lllrrrrrrl}    
    &    &     & frame & pixel& world &coord &     & counts per & nearby       \\
No. & RA & Dec &  x    &y & x & y &S/N & 1000sec    & counterpart  \\
\\

 1 & 22 40 30.05 & 3 21 28.94 & 304&299& 34.5& 14.5   &  14.4 &  6.2      &       Q2237+0305\\
 2 & 22 39 25.21 & 3 20 35.14 & 499&288& 1984.5&124.5 &  12.5 & 25.5    &   BD +02 4540,  V=9.6,                                     \\
   &             &            &    &   &       &      &       &         &                        G0-star (within fraction of arcsec) \\
 3 & 22 40  6.87 & 3 18 10.74 & 373&259& 724.5&414.5  &  10.1 &  2.9    &\\
 4 & 22 40 22.62 & 3 24 50.49 & 326&339&254.5&-385.5  &   5.2 &  1.1    &\\
 5 & 22 40 24.74 & 3 07 32.64 & 319&131&184.5&1694.5  &   4.6 &  2.8         &   HD 214787, V=8.3,                       \\
   &             &            &    &   &       &      &       &         &                        G5-star (50 arcsec off)\\
 6 & 22 40 50.78 & 3 23  8.02 & 242&318&-585.5&-175.5 &   7.3 &  1.7    &\\
 7 & 22 40 53.3  & 3 38 34.29 & 236&505&-645.5&-2045.5  &   4.5 &  9.8    &\\
 8 & 22 41 23.25 & 3 22 25.15 & 144&310&-1565.5&-95.5 &   3.6 &  1.1    &\\

\\
\end{tabular}
\end{center}
\caption{\label{tab-sources} 
Positions (J2000.0), signal-to-noise ratio, counts per ksec 
and comments on optical
identifications/nearby counterparts
for the eight sources detected in the field of quasar Q2237+0305.
The numbers correspond to the labels in Figure   \ref{fig-hri}.
}
\end{table*}

\begin{figure}[htb]
\psfig{figure=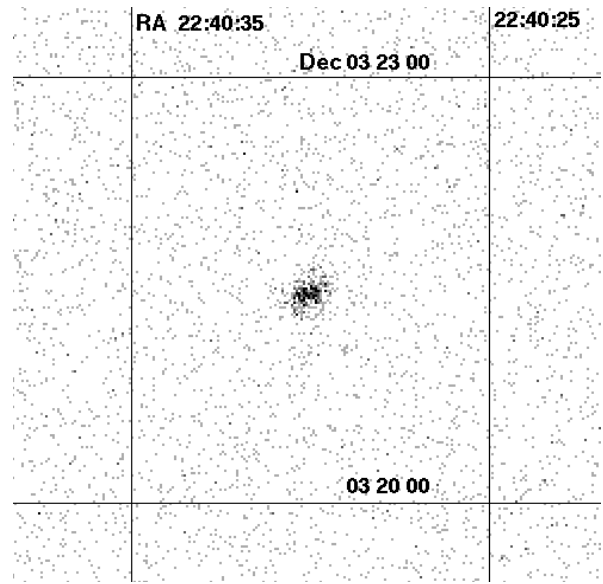    ,width=7.8cm,clip=} 
\caption{\label{fig-zoom} Zoom of the central part
of Fig. \ref{fig-hri} (here unsmoothed): 
ROSAT/HRI X-ray image of Q2237+0305; the four images are unresolved. 
}
\end{figure}

Figure \ref{fig-zoom} depicts a higher-resolution map of 
Q2237+0305. 
The image appears slightly elliptical; but the same small
ellipticity is seen in other images as well and hence seems
to be an artifact of imperfect pointing
(the separation of the four quasar images is only of order one
arcsecond; this cannot explain the apparent extension).

%
%
%
%
%

%
%
%
%
%
%

\section{Results and Discussion } \label{sec-results}

The average rate of $(6.2\pm2.8)$ counts/ksec 
translates into an energy flux of
$(2.2\pm1.0) \times 10^{-13}$ erg/sec/cm$^2$ in the 
interval 0.1-2.4 keV,
with the assumption of a hydrogen column density of $n_H = 5.5 \times
10^{20}$ cm$^{-2}$ (Dickey \& Lockman 1990) and a power law index 
(photon) of
$\alpha = 1.5$.  This flux can be converted into an X-ray luminosity
in the ROSAT energy window of $4.2 \times 10^{45}$erg/sec.  The
intrinsic X-ray luminosity of the quasar must be lower, since due to
the gravitational lensing there is a magnification of at least a few,
possibly even a few hundred (\cite{ken88,rix92,wam94}).

\begin{figure}[htb]
\psfig{figure=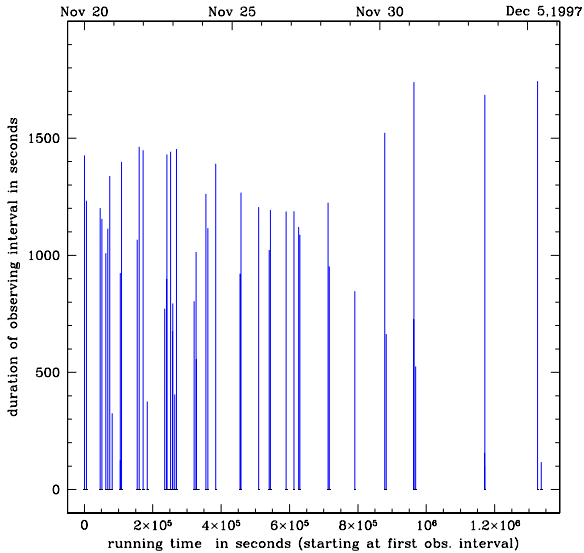       ,width=7.8cm,clip=} 
\caption{\label{fig-timing} 
Graphical ``log" of duration and epoch 
of the observing intervals for the ROSAT pointing
at Q2237+0305.
}
\end{figure}

As briefly mentioned in the introduction, it is of great interest to
study any variability of this quasar. The standard ROSAT/SASS analysis
did not find any indication for variability in Q2237+0305.  With only
about 330 source photons spread over two weeks in real time, it is
difficult to determine any variability.  In Figure \ref{fig-timing},
we display the observing intervals of this X-ray exposure for possible
comparison with observations in other wave bands at the same time.
The ``zero" on the time axis corresponds to November 20, 1997,
01:49:58 UT (or JD 2450772.576).  It is clear from this Figure that
the ``coverage factor" is less than 5\%.  So it makes no sense to
present a continuous light curve of Q2237+0305 over the observing
period.

%
%
%
%
%
%

\begin{figure}[htb]
\psfig{figure=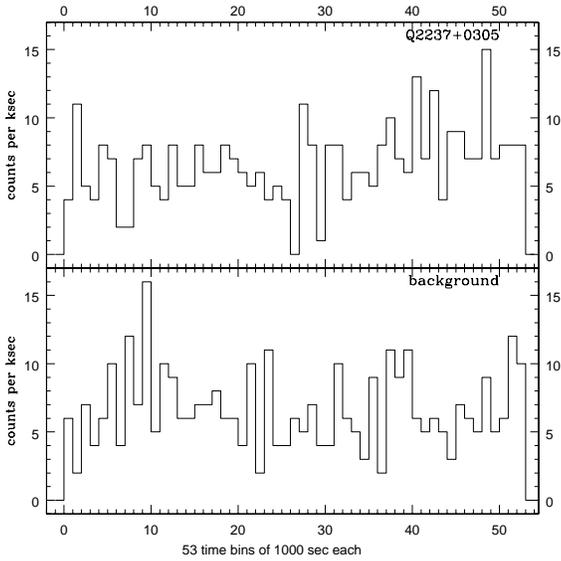      ,width=7.8cm,clip=} 
\caption{\label{fig-lightcurve} TOP: 
``X-ray light curve" for  Q2237+0305. 
This is the light curve as obtained by
attaching all observing intervals next to each other
(cf. Figure \ref{fig-timing}). 
Only eleven out of the 53 bins correspond to
real observing intervals with length 
1000 seconds or more. The photons were obtained in a circle
with a diameter of 15 arcsecond. 
The expected background contribution is 0.63 counts per 1000-sec-bin. 
The average count rate is 6.2 counts per ksec.
The abscissa gives the bin number. The bins are
1000 second wide.
BOTTOM: X-ray light curve for a background field with the same
average count rate as the above quasar light curve. 
The binning is identical for the top and bottom panels.
}
\end{figure}

We can also ``bin" the data to compare the Q2237+0305 ``light curve"
with the light curve of the ``background" (or of the other seven
detected sources, cf. Table \ref{tab-sources}).  Such a binned
artificial light curve is displayed in Figure \ref{fig-lightcurve} for
	bin widths of 1000 seconds\footnote{For the binning
	we basically put together all observing intervals back-to-back,
	thus ignoring all the ``dead times". The data in these
	53 bins of 1000sec each are in fact spread out over about 
	about two weeks (cf. Figure \ref{fig-timing})}.  
The top panel is the X-ray light curve
for Q2237+0305, the bottom panel is the ``light curve" 
of a background field with the same
average count rate (6.2 counts per ksec).
In order to find out whether the ``peaks" in the quasar light curve
could be due to enhanced background radiation, we chose 
to compare the quasar light curve  with  a background light curve
normalized to the same average count rate.
Figure \ref{fig-lightcurve} shows that the fluctuations in the 
flux of Q2237+0305
are uncorrelated with the variations in the background flux.

%
%
%
%
%
%

There is no obvious variability visible at the top of Figure
\ref{fig-lightcurve} that exceeds any variability in the bottom panel
(which sets the ``noise" level). Neither is there any ``correlated"
variability obvious, which could be expected for a highly increased
background at certain phases of the observation (since the extraction
circle of the Q2237+0305 light curve is much smaller than that for the
background, this would be surprising).

We investigated the issue of variability more quantitatively.  We
performed a KS-test with the real arriving times of the source photons
of Q2237+0305 (extraction radius 30 arcseconds), comparing them with
the background light curve of a circular region corresponding to the
same total number of photons and for the total background. Both were
consistent with no variability.  Furthermore, we performed a
chi-square test with the binned data. Similarly, we found no
indication for variability.

Among the 53 bins, the bin with the highest count contains 15 photons,
and one bin is completely empty.  The Poissonian probability for
finding 15 counts for an average of 6.2 is only  about
$1.2 \times 10^{-3}$, the
Poissonian probability for finding 0 counts with the same average is
$2.0 \times 10^{-3}$.

Similarly, if one divides the bins in two sets, the first 26 bins
contain 150 counts; the standard variation for an average of 150
counts is $\sigma_{26{\rm ksec}} = 12.25$ counts.  The second 26 bins
contain 194 counts, which is 2.4 $\sigma_{26{\rm ksec}}$ above the
counts of the first half.  These two tests leave open the possibility
of source variability, but at very low significance.

Another issue is a possible contamination of the X-ray counts by the
(lensing) galaxy. We estimated the X-ray luminosity of this galaxy
both by using Dell'Antonio et al. (1994)'s correlation $L_X(spiral) =
2.0 \times 10^{29} L_B^{1.0}$ (where $L_X$ is in erg s$^{-1}$ and
$L_B$ is in solar luminosities), as well as by extrapolating from the
known X-ray count rate of M31 (West et al. 1997). Both these estimates
result in a possible contribution of the lensing galaxy of less than
one percent of our detection, which hence can be neglected.

\section{Conclusions and Outlook}\label{sec-dis}

The detection of the quadruply-imaged quasar Q2237+0305 in X-rays with
($6.2\pm2.8$) counts per ksec opens up the possibility of being able to monitor
this system with the next generation of X-ray telescopes (Chandra
X-ray Observatory, XMM, ASTRO-E).  It would then be feasible to study
both the intrinsic variability of the quasar and microlens-induced
fluctuations.  The Chandra X-ray Observatory with its on-axis
resolution of 0.5 arcsec and its effective area almost
twice as large as ROSAT's 
will be able to detect and resolve the four images of
Q2237+0305.  In addition to the possibility of determining
microlens-induced fluctuations (see Yonehara et al. 1998), such
observations could offer the opportunity of measuring relative time
delays in this system.  Intrinsic X-ray variations of the lensed
quasar on time scales of less than a day would be required.  On the
other hand, if one could follow a ``caustic crossing event" in X-rays
(which should appear in only one of the four images, according to the
very high magnifications expected due to the small source size), we
could also have the possibility of determining the size or even the
source profile of the X-ray emission region of the quasar.

\begin{acknowledgements}
It is a pleasure to thank Ingo Lehmann for providing help with his
MIDAS tools.  This research has made use of the NASA/IPAC
Extragalactic Database (NED) which is operated by the Jet Propulsion
Laboratory, California Institute of Technology, under contract with
the National Aeronautics and Space Administration.
This research has also made use of the SIMBAD database, 
operated at CDS, Strasbourg, France.
\end{acknowledgements}

\end{document}